# Relation entre une cendre volante silico-alumineuse et son charbon

## P. ADAMIEC, J.C. BENEZET, A. BENHASSAINE

**Ecole des Mines d'Alès 6, Avenue de Clavières 30319 ALES Cedex**, e-mail:
Jean-Charles.Benezet@ema.fr

**Abstract:**

*The fly-ashes are typical complex solids which incorporate at the same time intrinsic properties with the layers (spectra mineralogical and dimensional spectra varied) and major transformations generated by the processes of development. To use fly-ashes in various applications, it is initially necessary to carry out a complete characterization of those. The first research to date carried out on the silico-aluminous fly-ashes in order to characterize them from the point of view physical, morphological, chemical and mineralogical resulted in saying that they are materials of a relative simplicity. To make this study, a silico-aluminous fly ash coming from the power station of Albi was selected. Heat treatments (450°C and 1200°C) made it possible to simulate the treatment undergone by coal in the power stations in order to be able to identify the residues. The diversity of the particles contained in ash could be explained by the relation existing between a fly ash and its coal of origin.*

**Pacs # : 05-70.-a, S 06, 07.85.Jy, S 81.05.Rm, 81.20.Ev, 81.30.Dz**

## 1. Introduction

Les cendres volantes sont issues des centrales thermiques et sont couramment utilisées dans l'industrie des ciments et bétons [1], des céramiques [2]. Les cendres volantes résultent d'une longue arborescence typique des procédés et des propriétés complexes. La figure 1 présente cette arborescence. La propriété convoitée est la production d'énergie électrique et les cendres volantes restent essentiellement un co-produit qu'il s'agit de valoriser.

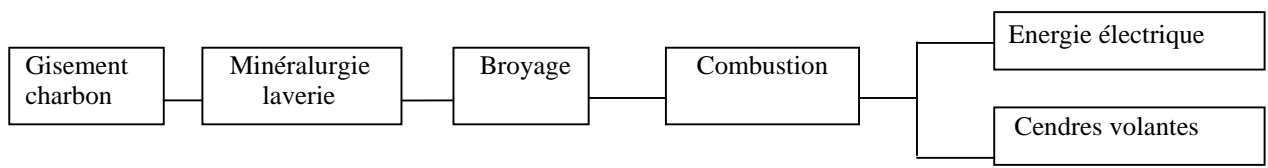

*Fig. 1 : Procédés de traitement du charbon conduisant à la formation de cendres volantes*

La cendre volante est aussi un solide complexe typique qui agrège à la fois des propriétés intrinsèques aux gisements (spectres minéralogiques et spectres dimensionnels variés) et des transformations profondes engendrées par l'arborescence. Pour l'utilisation des cendres volantes dans différentes applications, il est d'abord nécessaire de réaliser une caractérisation complète de celles-ci [3-5].

Les premières recherches effectuées à ce jour sur les cendres volantes silico-alumineuses en vue de les caractériser du point de vue physique, morphologique, chimique et minéralogique conduisent à dire que ce sont des matériaux d'une relative simplicité [6-11].

Simplicité du point de vue morphologique: les particules de cendres sont généralement présentées comme étant constituées en majorité de particules sphériques (figure 2).

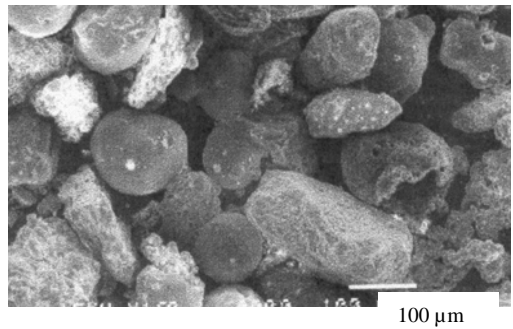

100 µm

*Fig. 2: Photo de la cendre volante*

Simplicité du point de vue chimique : la silice et l'alumine sont les deux éléments majeurs qui constituent la cendre. Cette population silico-alumineuse portée dans le diagramme ternaire (figure 3), pourrait pratiquement être associée à un diagramme binaire silico-alumineux.

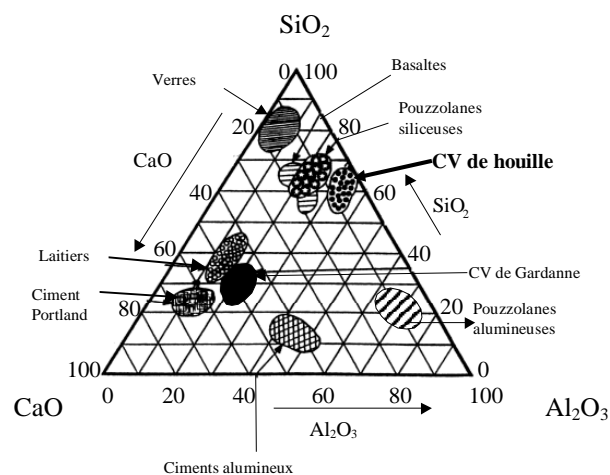

*Fig. 3 : Diagramme ternaire de KEIL-RANKIN*

Simplicité du point de vue minéralogique (figure 4): il existe une phase amorphe unique réactive (verre silico-alumineux) souvent prépondérante et des phases cristallisées inertes (mullite, quartz et magnétite).

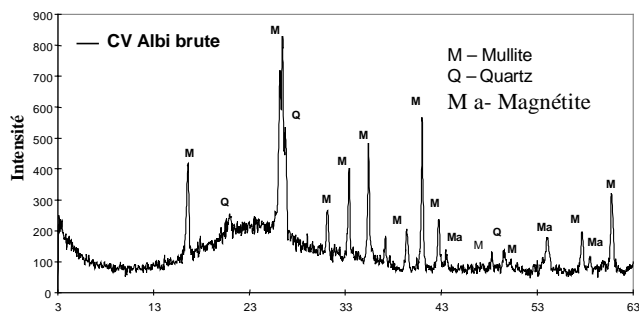

*Fig. 4 : Diffractogramme de la cendre volante brute d'Albi.*

De nombreux auteurs [6, 7, 12], ont doté les cendres volantes de propriétés physiques, chimiques, minéralogiques et morphologiques expliquées par un processus unique de transformation des particules au sein de la chaudière à travers un traitement thermique et des processus hydrodynamiques d'homogénéisation. Les cendres volantes ont été ainsi présentées comme un ensemble de particules de forme sphérique, de composition chimique simple, réduite à trois éléments (silice, alumine, et oxyde ferrique). Trois phases cristallisées (mullite, quartz et magnétite) et une phase amorphe unique (réactive ?) étaient identifiées.

Or, des études morphologiques et minéralogiques de plusieurs cendres volantes ont permis à leurs auteurs de proposer un modèle de classification des cendres volantes. Ces travaux de recherche montrent que les cendres volantes peuvent être considérées comme un mélange de plusieurs populations d'individus particuliers possédant des spécificités particulières et des origines diversifiées qui rendent obsolètes leurs mécanismes connus de formation. Le modèle le plus ancien [13] avait associé cendres volantes et charbons. Les minéraux répertoriés dans les charbons par ces auteurs relevaient tous du système $SiO_2$-$Al_2O_3$-$K_2O$. D'autres mécanismes de formation des cendres sont basés sur le modèle de fusion, expansion, sphéronisation et trempe [14, 15]. Le modèle de Hemming [14] accorde plus d'importance à la teneur en calcium et aux températures de fusion des eutectiques ternaires et à la viscosité de deux populations de cendres volantes.

## 2. Matériaux et méthodes

Pour vérifier la variété des cendres, nous avons cherché à relier la diversité d'une cendre volante à son charbon d'origine. Pour réaliser cette étude, une cendre volante silico-alumineuse provenant de la centrale thermique d'Albi a été choisie.

Une première calcination du charbon a été réalisée en lit fixe, à 450°C, pendant 10 heures. Ce traitement a permis d'éliminer la fraction carbonée et de récupérer les minéraux présents sans les modifier de façon significative.

Un second traitement thermique a fait suite à la calcination du charbon à 450°C. Ce traitement est constitué d'une montée en température (10°C/mn) jusqu'à 1200°C, puis d'un palier pendant 8 heures, et enfin d'un refroidissement jusqu'à la température ambiante (20°C/mn).

## 3. Fractionnement et caractérisation de la cendre

### 3.1. Fractionnement de la cendre

Conformément aux procédés minéralurgiques, un diagramme de traitement comportant trois opérations unitaires étagées a été établi (figure 5).

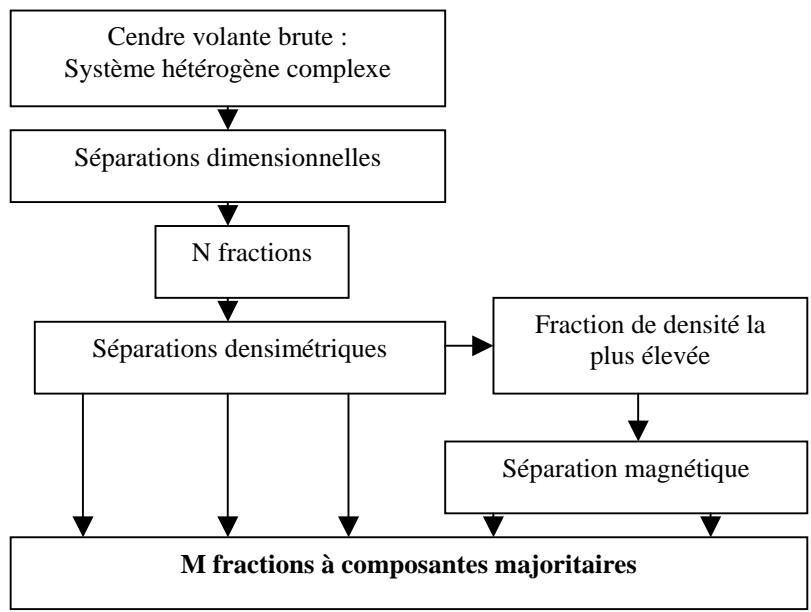

*Fig. 5 : Synoptique de traitement de la cendre*

Le premier étage a été consacré au fractionnement dimensionnel et a permis d'obtenir les produits notés A, B, C et D. Le second étage a permis de réaliser des séparations densimétriques sur chacune des fractions. Les densités utilisées ont été choisies à priori de manière à séparer les constituants (tableau 1).

| Densités | De 0 à 1,2 | De 1,2 à 1,8 | De 1,8 à 2,9 | > 2,9 |
|---|---|---|---|---|
| Nature des particules | Particules légères | Imbrûlés | Particules silico-alumineuses, quartz, verre | Mullite, particules lourdes d'oxydes de fer, corindon |

*Table 1 : Fractionnement densimétrique*

Le troisième étage est une séparation magnétique qui permet de retirer la classe de la densité la plus élevée (> 2,9) et d'enrichir cette fraction en composés riches en alumine : mullite et corindon.

Ce fractionnement a permis d'obtenir vingt classes (figure 6) qui sont répertoriées de la manière suivante :
- classes dimensionnelles indexées : A, B, C et D
- classes densimétriques indexées : 2, 3, 4 et 5
- classe magnétique indexée : 1.

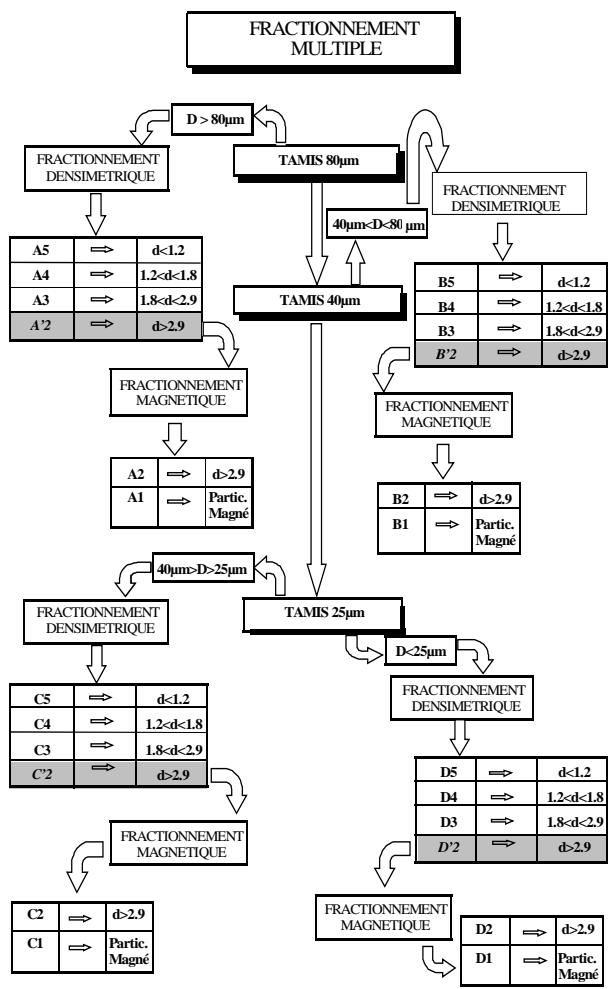

*Fig. 6 : Fractionnement appliquée à la cendre volante*

Les onze populations, qui représentées le plus d'importance du point de vue massique, ont été caractérisées.

## *3.2. Caractérisation des fractions*

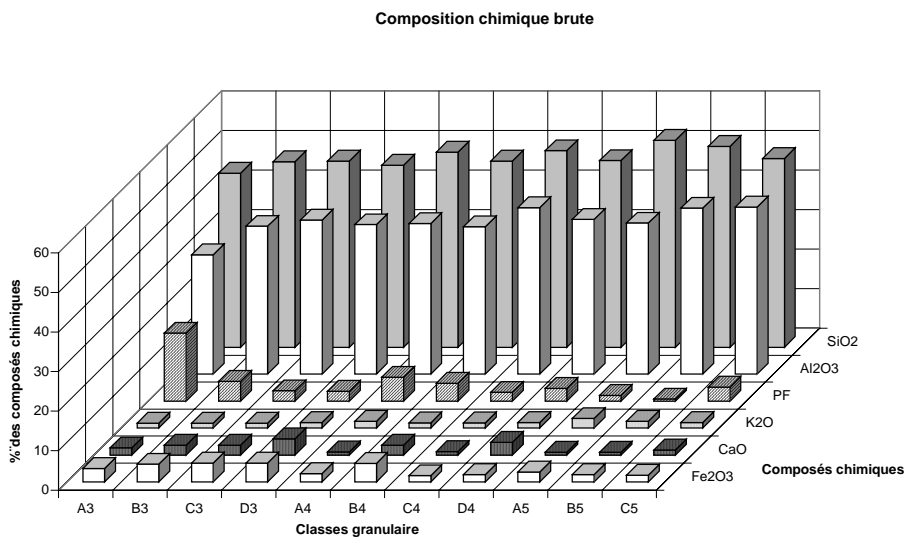

*Fig. 7 : Composition chimique des fractions de la cendre (PF : perte au feu)*

La composition chimique présente une relative constance dans les différentes fractions. La silice et l'alumine sont les constituants majoritaires tandis que le potassium, le calcium et le fer apparaissent en tant qu'éléments minoritaires (figure 7).

Les phases cristallisées dans la cendre globale et dans les sous classes sont identiques. Il s'agit de la mullite, du quartz et de la magnétite.

Le spectre densimétrique est très étendu. Les valeurs des densités varient entre 1 et 4,8 (figure 8).

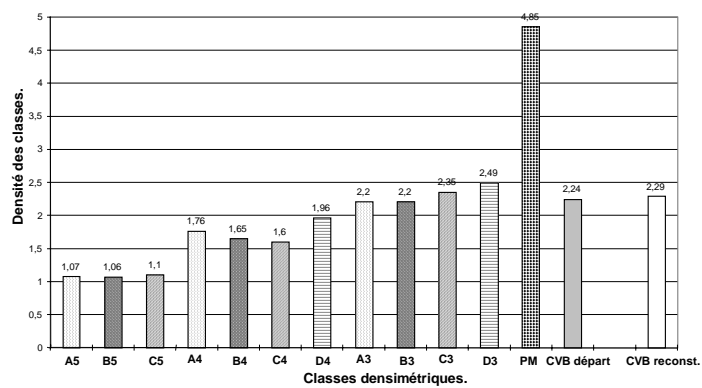

*Fig. 8 : Densité des classes de particules (PM : particules magnétiques)*

L'étude de la morphologie des particules, couplée à des analyses chimiques locales, peut permettre d'obtenir la typologie détaillée de la cendre volante d'Albi [16]. Différents types de particules ont été trouvés dans chaque fraction : des particules sphériques pleines, des sphères creuses, des grains alvéolaires, des grains de corindon, des grains de quartz, des particules magnétiques et des particules imbrûlées.

L'analyse chimique locale des particules a montré des compositions qui relevaient la plupart du temps du système binaire $SiO_2$-$Al_2O_3$ ou des ternaires simples. L'ensemble des résultats peut être symbolisé sur un diagramme comportant des systèmes ternaires multiples, qui permet de représenter les phases cristallisées (quartz, mullite, magnétite, hématite, chaux libre, anhydrite, corindon). Les phases amorphes présentent une très grande variation de composition. Elles appartiennent soit au système binaire $SiO_2$-$Al_2O_3$, soit aux ternaires simples et sont figurées dans le domaine hachuré (figure 9).

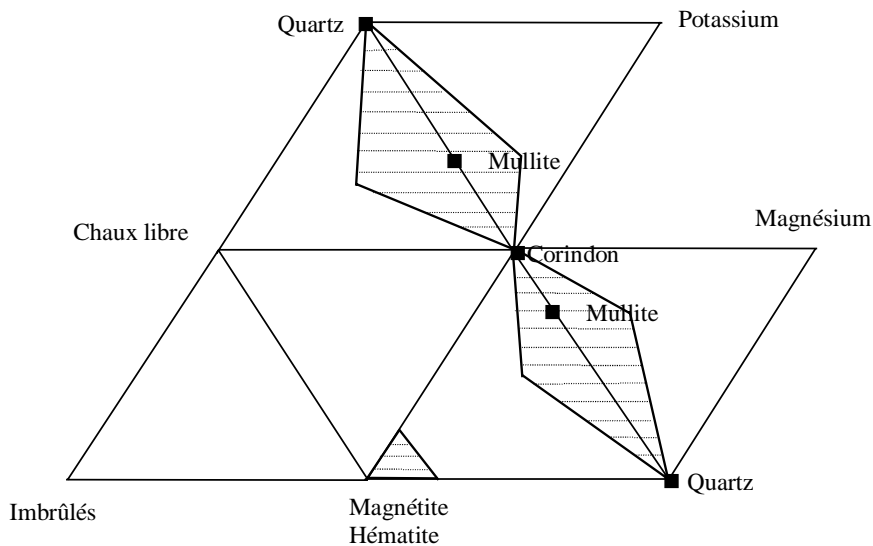

**Fig. 9 :** *Phases amorphes et cristallisées de la cendre d'Albi*

Cette diversité de compositions chimiques est totalement masquée par la prépondérance de la silice et l'alumine mais aussi par la représentation traditionnelle des cendres volantes dans le système de Keil-Rankin (figure 3).

## 4. Etude du charbon

Une large partie de la variabilité, souvent attribuée à l'échantillonnage, au fonctionnement de la centrale, pourrait résulter de la variété des cortèges minéraux dilués dans la phase carbonée. Une grande partie de la variabilité pourrait être héritée. L'étude du cortège minéral issu du charbon a été réalisée en trois étapes afin de simuler les étapes de traitement dans la centrale : étude du charbon non traité, du charbon traité à 450°C puis à 1200°C.

### 4.1. Etude du charbon non traité thermiquement

La quasi-totalité des types tomo-morphologiques identifiés dans les cendres volantes a été retrouvée, comme par exemple les particules sphériques pleines (figure 10) et creuses (figure 11).

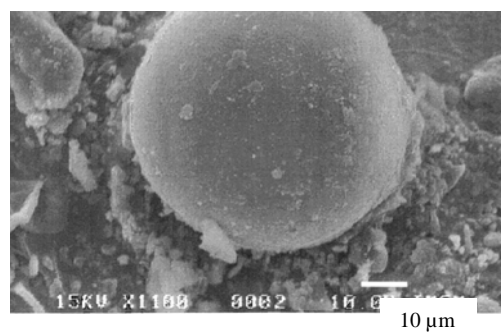

**Fig. 10 :** *Particules sphériques pleines*

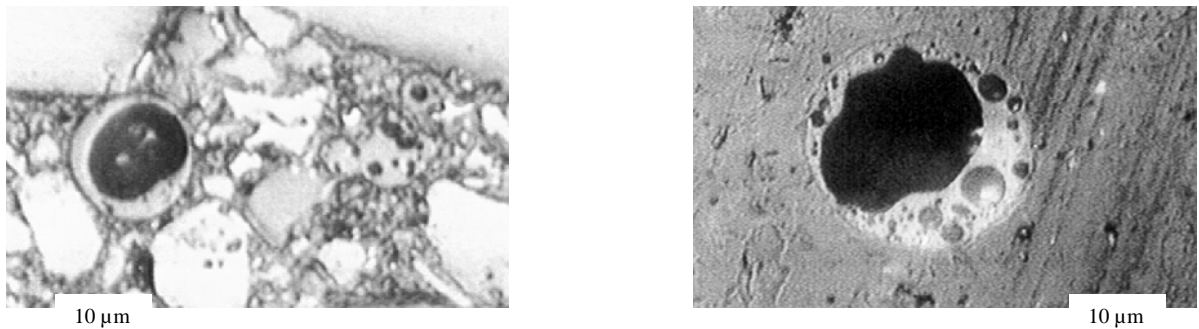

*Fig. 11 : Particules sphériques creuses*

Les analyses semi-quantitatives de ces particules sphériques creuses montrent que leur composition est semblable à celle des sphères creuses des cendres volantes: Si, Al, K et peu de fer.

Ces observations ne sont pas spécifiques au charbon d'Albi. Des examens similaires effectués sur d'autres charbons montrent l'existence de ces particules [17]. On retrouve ainsi dans le charbon non traité thermiquement des particules identiques à celles observées dans les cendres volantes.

### 4.2. Etude du charbon traité à 450°C

La kaolinite est le minéral le plus souvent cité comme présent dans les charbons. C'est aussi le minéral le plus facilement altéré par la température, sa transformation débute à une température avoisinant 490°C.

La forme de certaines particules est parfois altérée, notamment les sphères creuses dont les gaz occlus s'expansent jusqu'à la destruction des particules [18].
A 450°C, le cortège minéral est composé de quartz, de muscovite mal cristallisée (ou d'illite), de kaolinite et d'un dôme vitreux (figure 11).

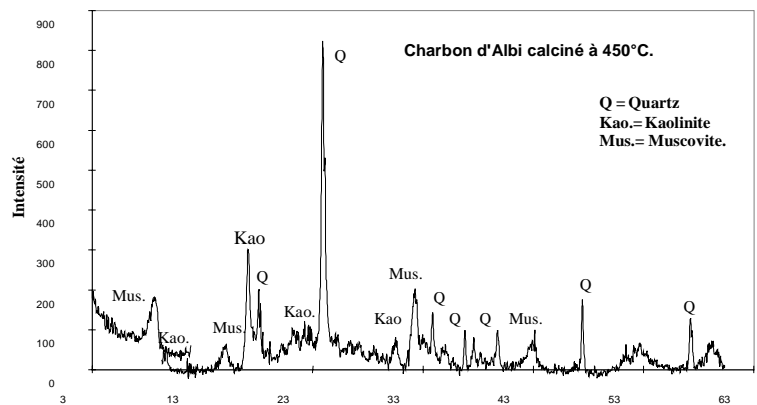

*Fig. 11 : Diffractogramme du charbon d'Albi traité à 450°C*

Ce cortège appartient en grande partie au système $SiO_2$-$Al_2O_3$-$K_2O$ et au système Carbone-Pyrite-Calcite, représentés sur la figure 12.

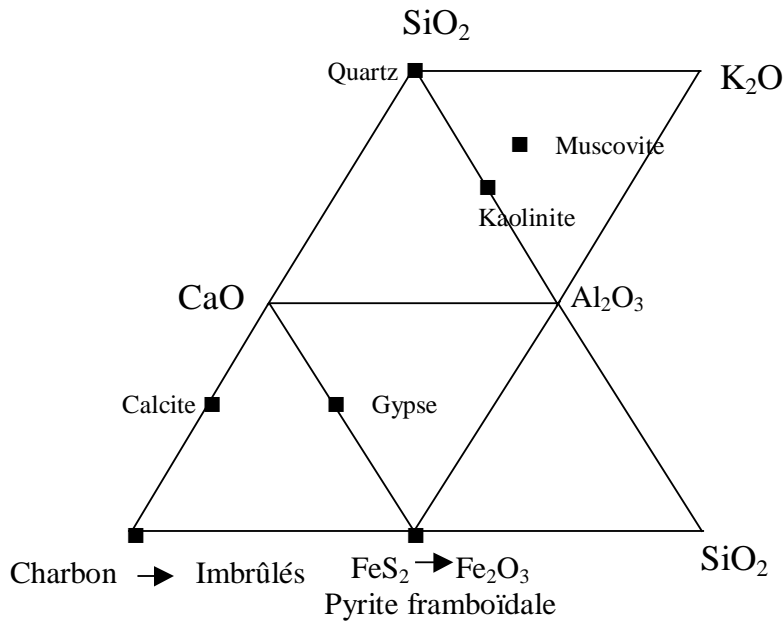

*Fig. 12 : Phases amorphes et cristallisées après traitement du charbon à 450°C*

### 4.3. Etude du charbon traité à 1200°C

Le résidu de ce traitement présente un aspect différent de celui d'une cendre volante. La couleur est rougeâtre, mais les observations microscopiques montrent que l'on y retrouve des particules semblables:

- les grains de quartz sont fracturés par l'effet thermique, les argiles se déshydroxylent, s'expansent et montrent des cristallisations de mullite,
- des particules magnétiques de formes arrondies, voire sphériques sont aussi présentes dans cette « cendre »,
- associée à la magnétite sous forme lamellaire, l'hématite se retrouve dans tous les grains,
- des sphères creuses de taille variable (entre quelques microns et 100 microns) ont été observées dans cette cendre.

Les caractérisations par diffraction des rayons X du charbon traité à 1200°C montrent deux phases cristallisées: le quartz en faible quantité et la mullite en quantité importante. Dans le même temps, un dôme vitreux apparaît (figure 13).

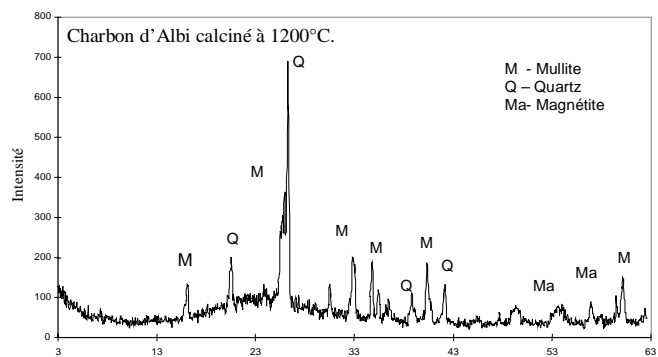

*Fig. 13 : Diffractogramme du charbon d'Albi traité à 1200°C*

A 1200°C, le cortège minéralogique s'est transformé : la muscovite a disparu et une nouvelle phase a cristallisé. Les phases, qui ont disparu, ont laissé la place à une phase désordonnée: le « verre ». Il est nécessaire d'identifier, de localiser ce « verre » puis de le relier à la kaolinite et à la muscovite (ou à l'illite). La cendre volante semble relever alors du binaire simple $SiO_2$-$Al_2O_3$.

Pour le charbon d'Albi, le cortège minéral du charbon constitué pour l'essentiel de l'association quartz-kaolinite-muscovite a été corrélé aux constituants minéralogiques d'une cendre volante constituée pour l'essentiel de l'association quartz-mullite-verre (figure 14).

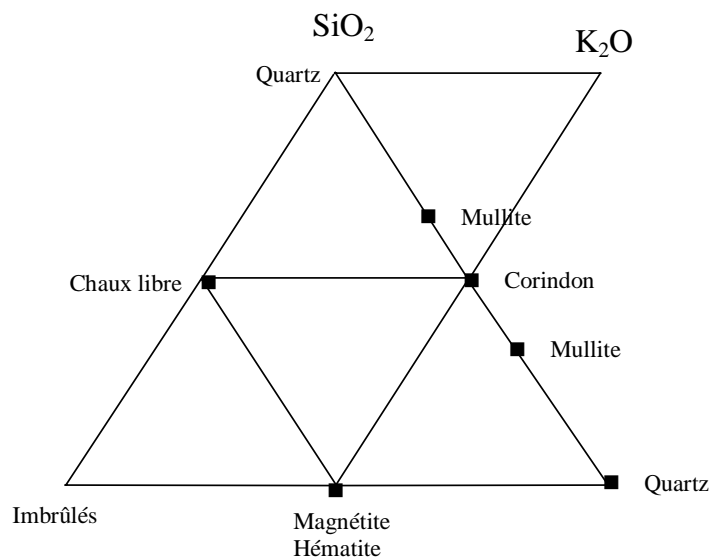

*Fig. 14: Phases amorphes et cristallisées après traitement à 1200°C*

## 5. Conclusions

La première conclusion de ce travail fait apparaître clairement que des valeurs moyennes macroscopiques physiques et chimiques peuvent cacher l'existence de particules de compositions chimiques très variées ainsi qu'une grande diversité morphologique.

Au cours de cette étude, la part d'héritage de la cendre volante d'Albi a été établie. S'il est possible de retrouver la composition chimique et minéralogique de la cendre volante, il n'est plus possible de parler de « taux de verre » ou même de « verre » comme phase unique.

Compte tenu de la largeur du domaine cristallographique, le dôme vitreux recouvre un large spectre de phases « désorganisées » et opacifie la lecture du système. La variabilité d'une cendre volante provient à la fois de la variété du cortège minéral et des quantités respectives de chaque minéral. L'existence d'un dôme commun des minéraux de compositions variées confirme que la notion de « verre silico-alumineux » reliée à la notion de phase unique risque de conduire à de véritables contresens en terme de réactivité. Il apparaît préférable de considérer plutôt un mélange de phases désorganisées. Bien évidemment le premier souci d'un utilisateur des cendres pour la réactivité [19] sera d'identifier l'origine de ces phases désorganisées.

Le traitement thermique sur le charbon d'Albi a permis de mettre en évidence le lien entre les cendres et leurs charbons d'origine :

- le quartz est un minéral hérité,
- la kaolinite conduit à la cristallisation de la mullite et de la cristobalite,
- la gibbsite conduit à la cristallisation du corindon,
- la plupart des silico-aluminates montre un effondrement des organisations minérales, qui se traduit par la manifestation d'un dôme vitreux.

## Références :